\documentclass[preprint,12pt]{elsarticle}

\usepackage{amssymb}
\usepackage{amsmath}
\usepackage[table]{xcolor}

\newcommand{\RVA}{\mathcal{R}_{V\!A}}
\newcommand{\om}{\omega}
\newcommand{\braket}[1]{\left\langle#1\right\rangle}
\newcommand{\mrm}[1]{\mathrm{#1}}
\newcommand{\psibar}{\overline{\psi}}
\newcommand{\order}{{\cal O}}

\newcolumntype{a}{>{\columncolor{green!50}}r}
\newcolumntype{b}{>{\columncolor{green!50}}c}
\graphicspath{{./}{figures/}}

\begin{document}

\begin{frontmatter}



\title{Approaching the continuum with anisotropic lattice thermodynamics}

\author[Maynooth,TCD]{Jon-Ivar Skullerud}
\author[Swansea]{Gert Aarts}
\author[Swansea]{Chris Allton}
\author[Swansea]{M. Naeem Anwar}
\author[TCD]{Ryan Bignell}
\author[Swansea]{Tim Burns}
\author[Liverpool]{Simon Hands}
\author[Maynooth]{Rachel Horohan D'Arcy}
\author[Odense]{Ben J\"ager}
\author[Sejong]{Seyong Kim}
\author[Swansea]{Alan Kirby}
\author[Firenze]{Maria Paola Lombardo}
\author[Busan]{Seung-Il Nam}
\author[TCD]{Sin\'ead M. Ryan}
\author[Swansea]{Antonio Smecca}


\affiliation[Maynooth]{organization={Department of Physics,
  Maynooth University---National University of Ireland Maynooth},
  city={Maynooth}, postcode={Co Kildare}, country={Ireland}}
\affiliation[TCD]{organization={School of Mathematics, Trinity College},
  city={Dublin}, country={Ireland}}
\affiliation[Swansea]{organization={Department of Physics, Swansea University},
       addressline={Singleton Park, Swansea SA2 8PP}, country={UK}}

\affiliation[Liverpool]{
  organization={Department of Mathematical Sciences, University of Liverpool},
  city={Liverpool}, postcode={L69 3BX}, country={UK}}
\affiliation[Odense]{
  organization={Quantum Field Theory Center \& Danish IAS,
  Department of Mathematics and Computer Science},
  addressline={University of Southern Denmark, 5230 Odense M},
  country={Denmark}}
\affiliation[Sejong]{
  organization={Department of Physics, Sejong University},
  city={Seoul}, postcode={05006}, country={Korea}
  }
\affiliation[Firenze]{
  organization={INFN, Sezione di Firenze},
  addressline={50019 Sesto Fiorentino}, postcode={(FI)},
  country={Italy}
}
\affiliation[Busan]{
  organization={Department of Physics, Pukyong National University (PKNU)},
  city={Busan}, postcode={48513}, country={Korea}
}

\begin{abstract}
The FASTSUM collaboration has a long-standing programme of using
an\-iso\-tro\-pic lattice QCD to investigate strong interaction
thermodynamics, and in particular spectral quantities.  Here we
present first results from our new ensemble which has a temporal
lattice spacing $a_\tau=15\,$am and anisotropy $\xi=a_s/a_\tau=7$, giving
unprecedented resolution in the temporal direction.  We show results
for the chiral transition, vector--axial-vector degeneracy, and heavy
quarkonium, and compare them with earlier results with coarser time resolution.
\end{abstract}





\end{frontmatter}



\section{Introduction}
\label{sec:intro}

Many of the most interesting, yet poorly determined properties of the
quark--gluon plasma (QGP) created in heavy-ion collisions are spectral
quantities.  These include, firstly, transport properties such as
conductivity, diffusion coefficients, and viscosities; and secondly,
hadronic properties such as their degeneracy patterns, mass shifts,
width, and dissociation in the QGP.  Determining such properties from
euclidean correlation functions is known to be an ill-posed problem as
it involves solving the integral function
\begin{equation}
  G_E(\tau;T) = \int_0^\infty d\om K(\om,\tau;T)\rho(\om;T)\,,
\end{equation}
where $G_E$ is the euclidean correlator, $\rho$ is the (unknown)
spectral function, and $K$ is a known integral kernel.  The ill-posed
nature of this problem arises from a combination of the limited number
of euclidean data points and their statistical uncertainties.  In
order to mitigate the first of these issues, the FASTSUM collaboration
has been using anisotropic lattices, where the temporal resolution is
much finer than the spatial one.  This has yielded a wealth of
information including on heavy quark physics \cite{Aarts:2007pk,Aarts:2011sm,Aarts:2014cda,Kelly:2018hsi,Aarts:2022krz,Aarts:2023nax,Skullerud:2025iqt}, electrical
conductivity \cite{Amato:2013naa,Aarts:2014nba}, and parity doubling in baryons \cite{Aarts:2015mma,Aarts:2017rrl,Aarts:2018glk,Aarts:2023nax}.

\begin{table}[t]
\begin{center}
\begin{tabular}{c|cllccrr}\hline
Gen & $N_f$& $\xi$  & $a_s$ (fm) & $a_\tau^{-1}$ (GeV) & $m_\pi$ (MeV) & $N_s$
& $L_s$ (fm)  \\ \hline
1 & 2   & 6.0 &  0.162(4) &   7.35(3) & 490 & 12 & 1.94 \\
2 & 2+1 & 3.45 &  0.1205(8) &   5.63(4) & 384(4) & 24 & 2.95 \\
 & & & & & & 32 & 3.94 \\
2L & 2+1 & 3.45 &  0.1121(3) &   6.08(1) & 239(1) & 32 & 3.58 \\
\rowcolor{green!50}
3 & 2+1 & 7.0 & 0.108(1) & 12.82(7) & 378(1) & 32 & 3.49 \\
\hline
\end{tabular}
\end{center}
\caption{FASTSUM ensembles: ensemble label (Generation), number of
  quark flavours $N_f$, spatial and temporal lattice spacings
  $a_s,a_\tau$ and anisotropy $\xi=a_s/a_\tau$, pion mass $m_\pi$ and
  spatial extent $L_s=a_sN_s$.}\label{tab:ensembles}
\end{table}

The parameters of the ensembles we have used to date are given in
table~\ref{tab:ensembles}.  In these proceedings, we present initial
results from our newest ``Generation 3'' ensemble, where the temporal
lattice spacing is about half that of our previous ensembles.  This
will give us unprecedented temporal resolution and hence assist
spectral reconstruction, while also allowing a denser set of
temperatures than hitherto possible in the fixed-scale approach.  We
present our procedure for tuning the parameters in
section~\ref{sec:params}.  In section~\ref{sec:chiral} we present
initial results for the chiral transition including the
vector--axial-vector degeneracy, while section~\ref{sec:beauty}
presents preliminary results for the temperature dependence of heavy
quarkonium.  We conclude with an overview of planned future work on
this and future ensembles.

\section{Tuning the action parameters}
\label{sec:params}

Our simulations are carried out using anisotropic lattices with an
$\order(a^2)$ improved gauge action and an $\order(a)$ improved Wilson
fermion action with stout smearing, following the approach of the
HadSpec collaboration \cite{Edwards:2008ja,Lin:2008pr}, who carried
out the tuning for our Gen2 and Gen2L ensembles.  The details of the
action are given in \cite{Aarts:2020vyb}.  It contains four
independent parameters: the gauge coupling $\beta$, the bare gauge and
fermion anisotropies $\gamma_g,\gamma_f$, and the bare quark mass
$m_0$. These must be tuned simultaneously to obtain the desired values
for the quantities $a_s,\xi_g,\xi_f,m_\pi$, where $\xi_g$ and $\xi_f$
are measured anisotropies in the gauge and fermion sector
respectively.

We have carried out the parameter tuning at the 3-flavour symmetric
point, with the target values
\begin{align}
 a_s &\approx a_s(\text{Gen2}) = 0.12\,\mrm{fm}\,,
 & \xi_g = \xi_f &= 7.0\,,
 &\frac{m_{PS}^{\text{3fl}}}{m_V^{\text{3fl}}} &= 0.545\,,
\label{eq:targets}
\end{align}
where the last value is obtained from the 3-flavour symmetric
equivalent of the Gen2 values,
\begin{align}
\frac{1}{3}\Big(\frac{m_{PS}^{\text{3fl}}}{m_V^{\text{3fl}}}\Big)^2
= \frac{m_\pi^2+2m_K^2}{(m_\rho+2m_{K^*})^2}\Big|_{\text{Gen2}}\,.
\end{align}
Following this, $\beta,\gamma_g$ and $\gamma_q$ have been fixed and
the light and strange bare quark masses tuned to reproduce the Gen2
pion mass, $m_\pi=390\,$MeV, and the physical value for
$m_{\eta_s}/m_\phi$, where $m_{\eta_s}=\sqrt{2m_K^2-m_\pi^2}$ is the
mass of the would-be $s\bar{s}$ pseudoscalar.

For the 3-flavour tuning, we have generated $\sim100$ ensembles of
$N_\tau\times N_s^3=256\times24^3$ lattices with
$\sim2000$ molecular dynamics trajectories each.

\subsection{3-flavour tuning}
\label{sec:Nf3-tuning}

\begin{figure}
\includegraphics[width=0.45\textwidth]{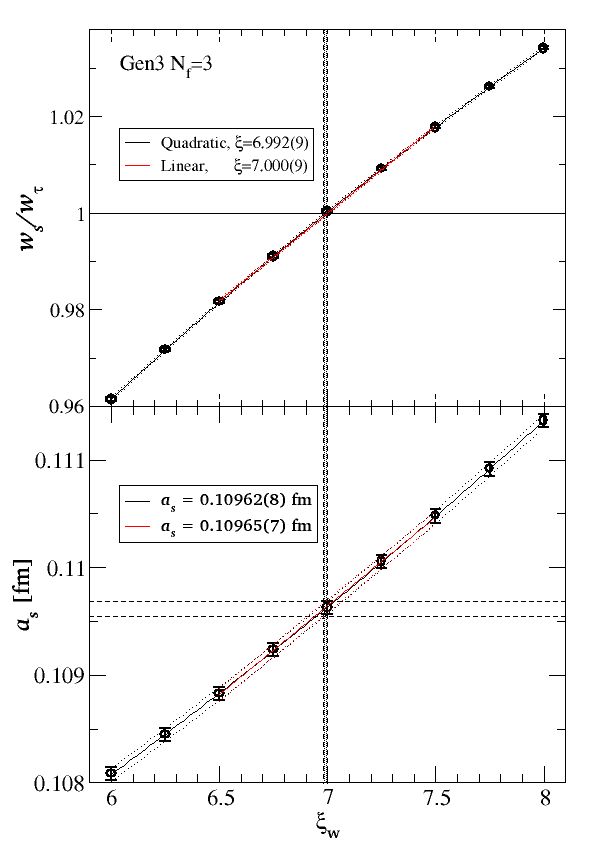}
\includegraphics[width=0.45\textwidth]{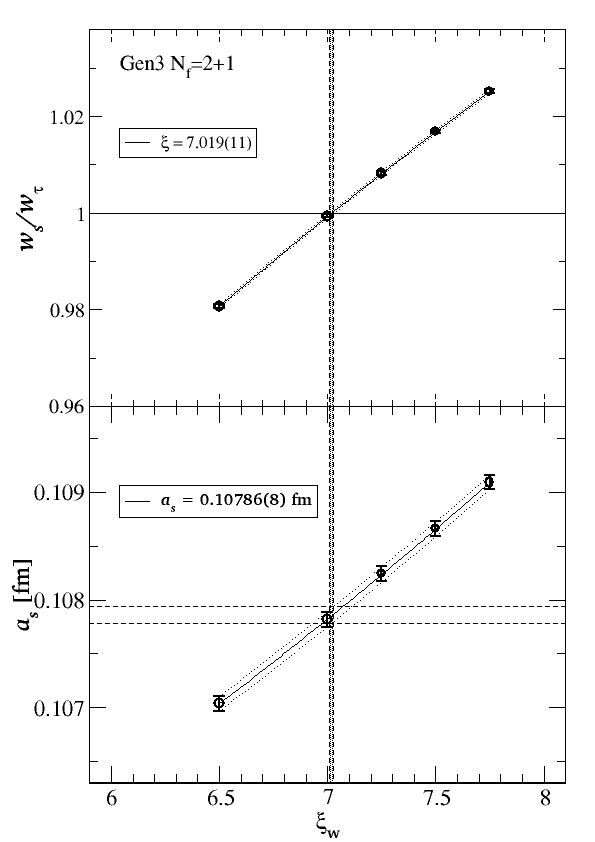}
\caption{Gauge anisotropy (top) and lattice spacing (bottom) from the
  Wilson flow.  Left: $N_f=3$, including linear and quadratic fits.
  Right: $N_f=2+1$, including quadratic fits.}
\label{fig:wflow}
\end{figure}

The gauge anisotropy and (spatial) lattice spacing have been
determined using the Symanzik flow method introduced in
\cite{Borsanyi:2012zr}.
For an isotropic lattice, the $w_0$ scale is defined as where the flowed
fields satisfy the condition
\begin{equation}
\Big[t\frac{d}{dt}t^2E(t)\Big]_{t=w_0^2} = 0.3\,,\qquad
E(t) = \frac{1}{4}\sum_{\mu\nu}F_{\mu\nu}^2\,,
\label{eq:w0-scale}
\end{equation}
and we take the most recent FLAG \cite{FLAG:2024oxs} value, $w_0=0.17355(92)\,$fm.
In the case of an anisotropic lattice, with $a_s = \xi a_\tau$, we have
\begin{equation} 
  E_{ij}^{\text{phys}} = a_s^4E_{ij}^{\text{lat}}\,,\qquad
  E_{i4}^{\text{phys}} = a_s^2a_\tau^2E_{i4}^{\text{lat}}\,,
  \label{eq:anisotropic-flow}
\end{equation} 
and the flow equation now includes the flow anisotropy parameter $\xi_w$.
We may also define the spatial and temporal $w$-scales, $w_s, w_\tau$:
\[
\Big[t\frac{d}{dt}t^2E_{ss}(t)\Big]_{t=w_s^2} = 0.15\,,\qquad
\xi_w^2\Big[t\frac{d}{dt}t^2E_{s\tau}(t)\Big]_{t=w_\tau^2} = 0.15\,.
\]
The physical gauge anisotropy $\xi_g$ can now be determined as the
value for $\xi_w$ where $w_s=w_\tau$, and the lattice spacing is
determined from the value of $w_s=w_\tau=w_0$ at this flow anisotropy.
Figure~\ref{fig:wflow} illustrates this for our final 3-flavour
symmetric ensemble as well as for the 2+1 flavour ensemble.  We see
that the anisotropy and lattice spacing are both almost unchanged when
going from $N_f=3$ to $N_f=2+1$ at fixed $\beta,\gamma_g,\gamma_f$.

\begin{figure}
\centering
\includegraphics[width=0.45\textwidth]{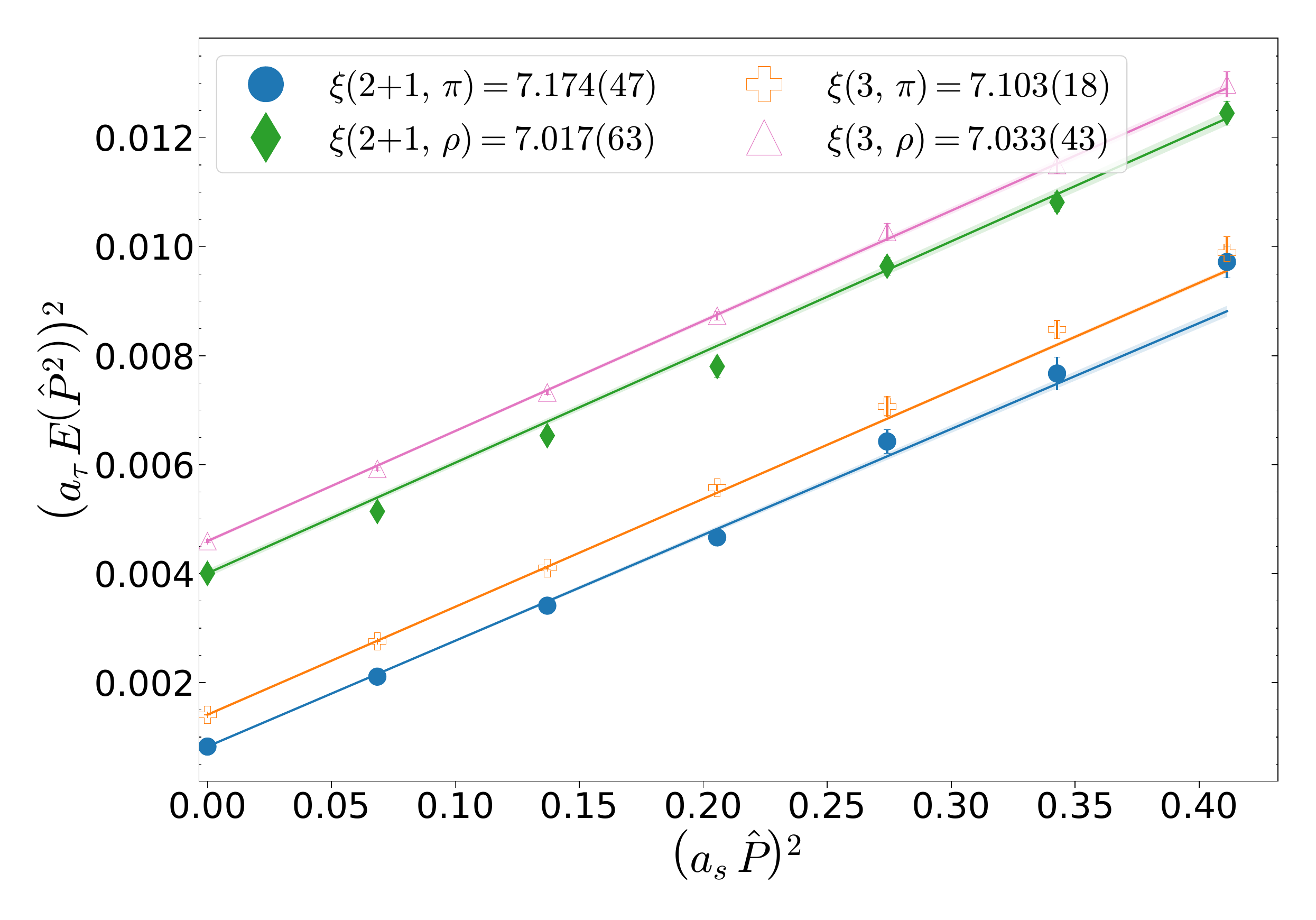}
\includegraphics[width=0.45\textwidth]{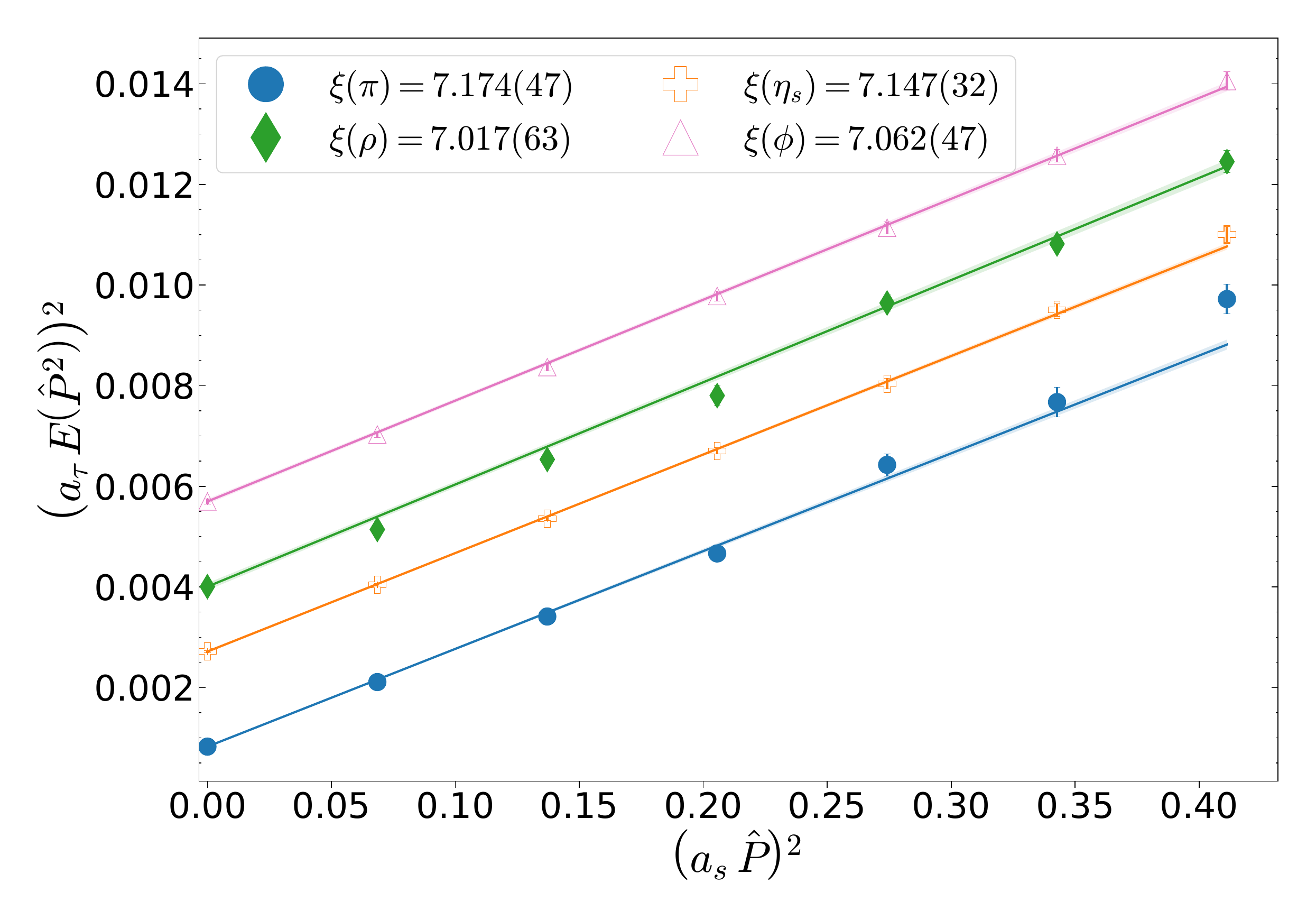}
\caption{Left: light pseudoscalar and vector dispersion relation for
  the $N_f=3$ and $N_f=2+1$ ensembles.  Right: Light and strange meson
  dispersion relations for the $N_f=2+1$ ensemble.  Note that the blue
circles and green diamonds are the same in both plots.}
\label{fig:dispersion}
\end{figure}

The fermion anisotropy is determined from the pseudoscalar and vector
meson dispersion
relations, as shown in 
Fig.~\ref{fig:dispersion}.  From the 3-flavour symmetric dispersion
relations in the left panel, we
see that the resulting anisotropies agree very well both with each
other and with the target (and gauge) anisotropy of 7.0.  The spatial
lattice spacing comes out slightly smaller than for the Gen2 ensemble,
which should be taken into account when comparing results from the
two.

\subsection{2+1 flavour tuning}
\label{sec:Nf2+1-tuning}

\begin{figure}
\centering
\includegraphics[width=0.45\textwidth]{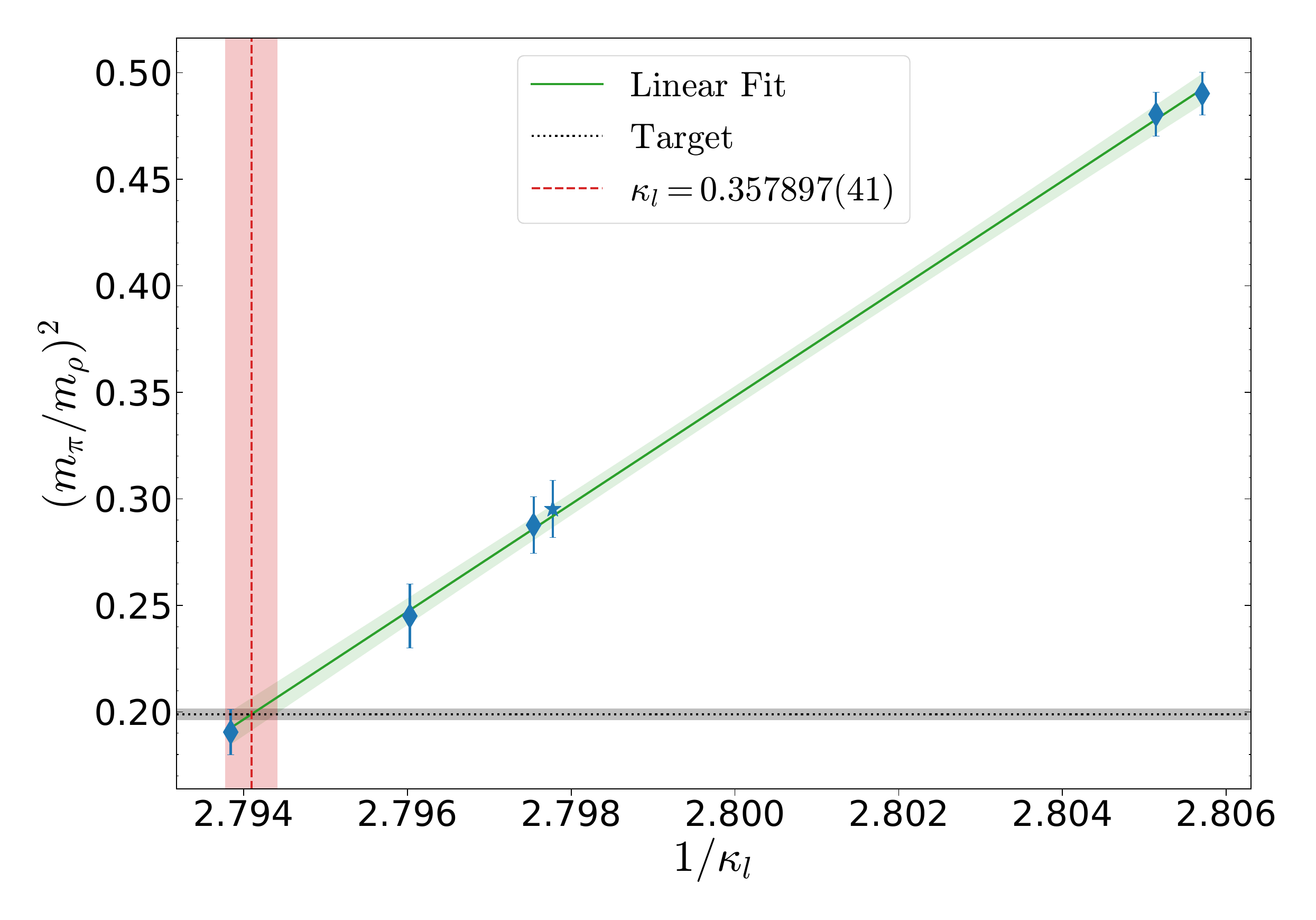}
\includegraphics[width=0.45\textwidth]{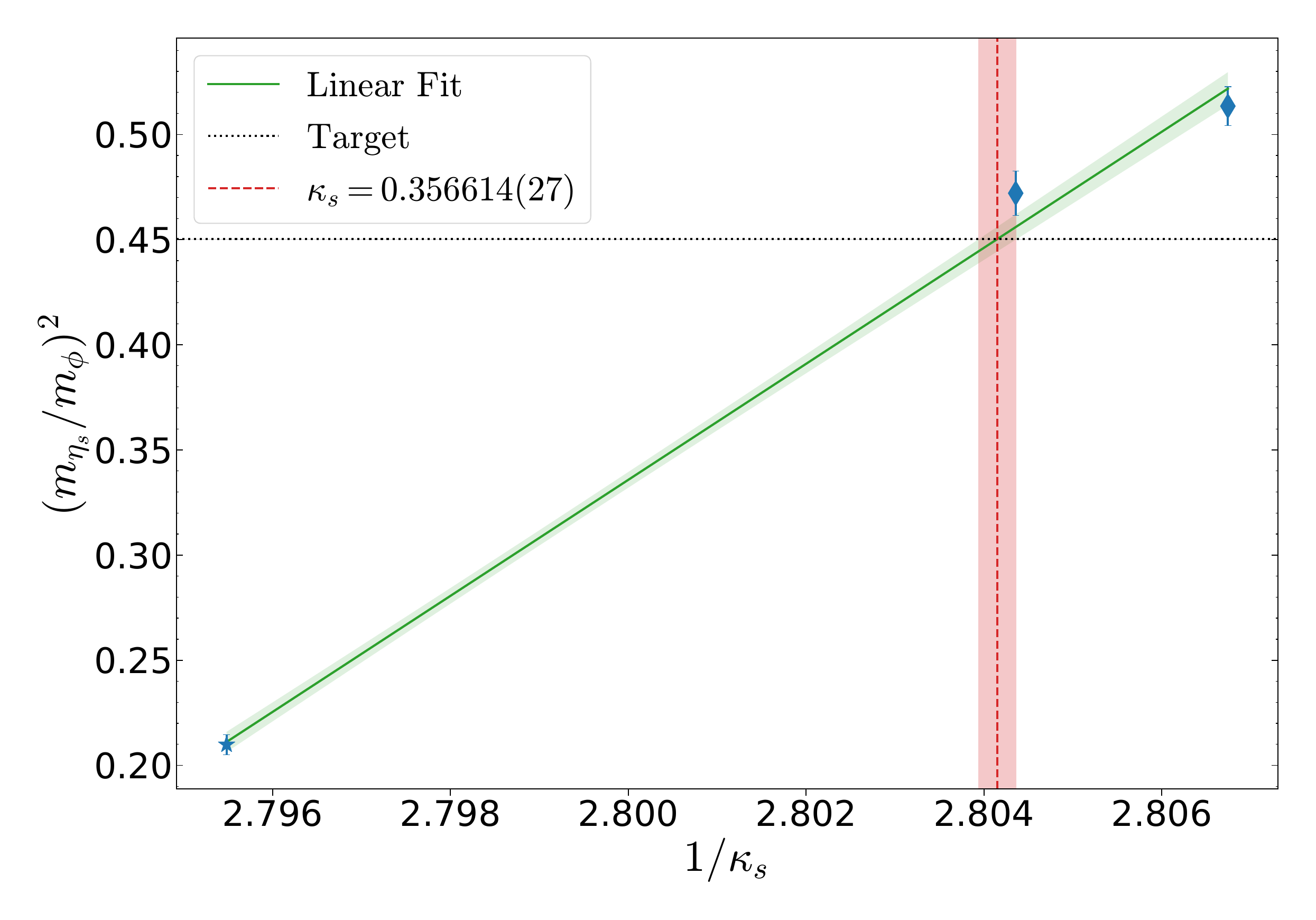}
\caption{Left: The squared pseudoscalar-to-vector meson mass ratio as a
  function of the bare quark mass (inverse hopping parameter), for a
  number of valence quark masses on the 3-flavour ensemble.  Right:
  The strange pseudoscalar-to-vector mass ratio on an $N_f=2+1$
  background.}\label{fig:mass-tuning}
\end{figure}

Following the 3-flavour tuning, the light quark mass was determined by
producing correlators with various valence quark masses on the
3-flavour ensemble, as shown in Fig.~\ref{fig:mass-tuning} (left).
This produced estimates for the light and strange bare quark masses
which were used to generate $N_f=2+1$ ensembles, and the procedure was
iterated to yield the final mass parameters.  The strange quark mass
was tuned by requiring that the dimensionless ratio
$m_{\eta_s}/m_\phi$ takes on its physical value, as shown in
Fig.~\ref{fig:mass-tuning} (right).  We have confirmed that different
methods to fix the strange quark mass give consistent results.

Finally, the lattice spacing and gauge and fermion anisotropies were
determined using the same procedures as in Sec.~\ref{sec:Nf3-tuning},
as shown in the right panels of Figs~\ref{fig:wflow} and
\ref{fig:dispersion}. This yields results consistent with those from
the 3-flavour ensembles, with the lattice spacing slightly reduced.
Furthermore, we find that the $\Omega^-$ 
mass comes out close to its physical value, or conversely, that
setting the scale from the $\Omega^-$ mass gives results consistent
with the $w_0$ scale setting.

\section{Results}
\label{sec:results}

Using our $N_f=2+1$ parameters from the tuning process, we have
generated gauge configurations at 14 temperatures ranging from 100 to
530 MeV, in addition to our zero-temperature ($256\times24^3$) ensemble.
At each temperature we have generated approximately 1000
configurations separated by 10 MD trajectories.  In the following we
present some preliminary results from these ensembles.


\subsection{Chiral transition}
\label{sec:chiral}

First we consider the standard order parameter of the chiral
transition, the chiral condensate $\braket{\psibar\psi}$, and its
derivative, the
chiral susceptibility $\chi_{\psibar\psi}$.
In the fixed-scale approach, the (additive and multiplicative)
renormalisation is temperature-independent, so for the purposes of
determining the chiral transition temperature it is sufficient to
consider unrenormalised (but vacuum subtracted) quantities, which is
what we will do here.

\begin{figure}
\centering
\includegraphics[width=0.45\textwidth]{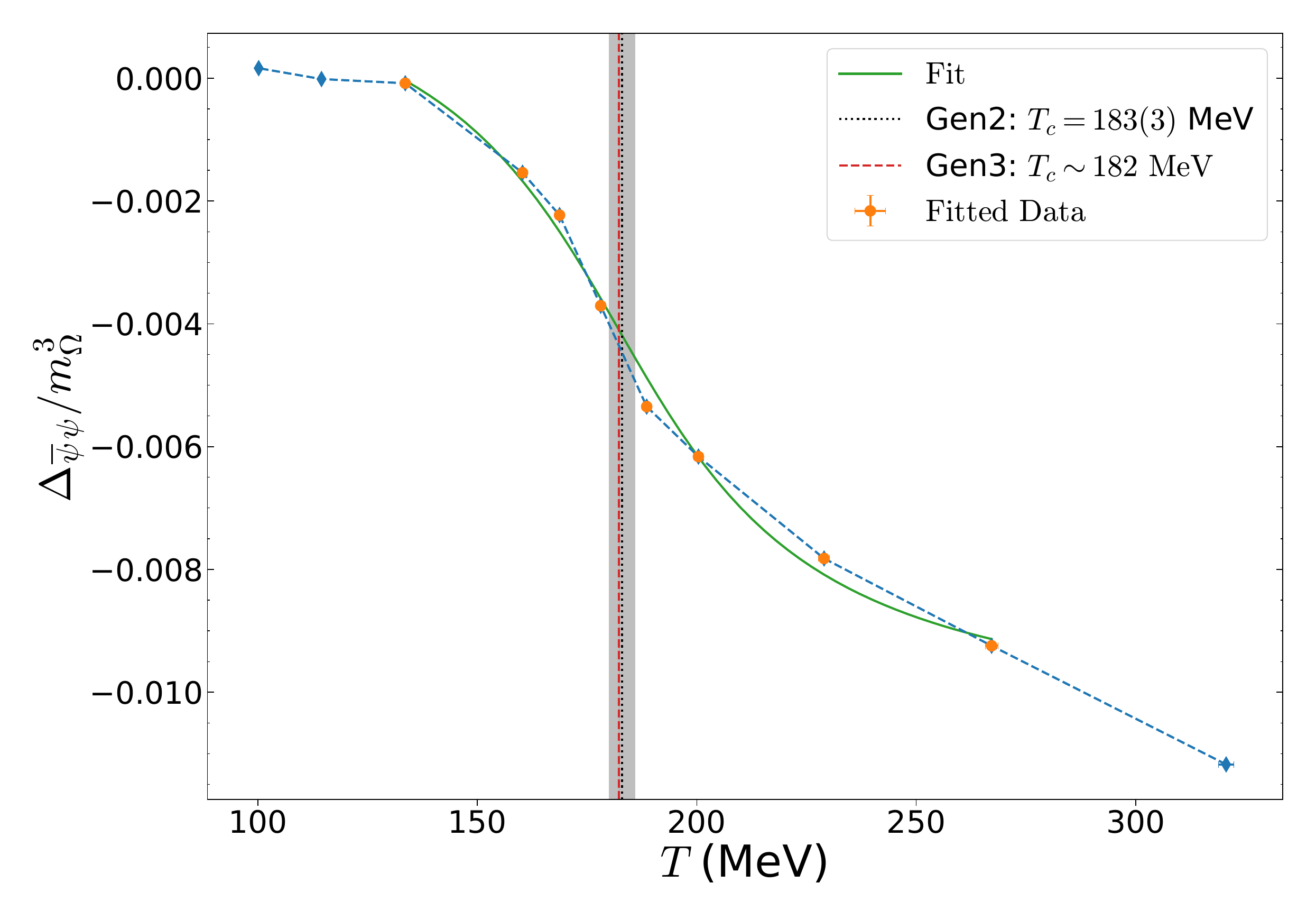}
\includegraphics[width=0.45\textwidth]{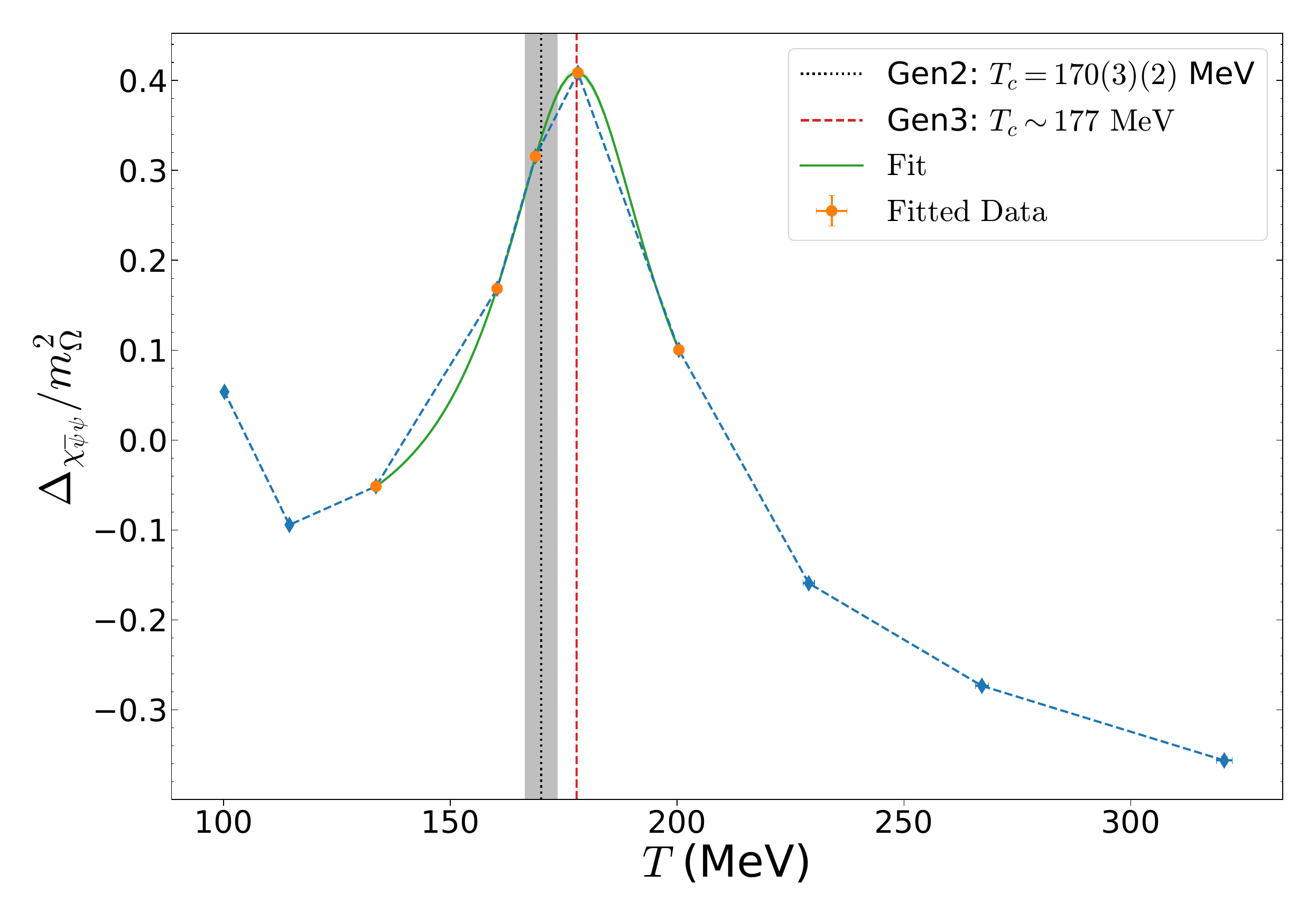}
\caption{Left: The dimensionless bare subtracted chiral
  condensate. Right: The
  dimensionless chiral susceptibility.  The dashed lines are there to
  guide the eye, while the continuous lines are fits to
  \eqref{eq:transition-fits}, using the data points represented by the
orange diamonds.}
\label{fig:chiral}
\end{figure}

In Fig.~\ref{fig:chiral} we show the bare, vacuum subtracted chiral
condensate $\Delta_{\psibar\psi}=\braket{\psibar\psi}-\braket{\psibar\psi}_0$ and
susceptibility $\Delta_\chi=\chi_{\psibar\psi}-\chi_{\psibar\psi,0}$.  These
quantities have been made dimensionless by dividing by appropriate
powers of the $\Omega^-$ mass.  The pseudocritical temperature is
given by the inflection point in $\Delta_{\psibar\psi}$ or the peak in
$\Delta_\chi$; we determine this by fitting the data to functions
given by
\begin{align}
  \Delta_{\psibar\psi} &= c_1 + c_2\arctan\big[c_2(T-T_c^{\psibar\psi})\big]\,;
  &\Delta_{\chi} &= c_2 + \frac{c_0}{c_1 + (T - T_c^{\chi})^2}\,.
  \label{eq:transition-fits}
\end{align}
This gives us $T_c^{\psibar\psi}\sim182\,$MeV and
$T_c^{\chi}\sim178\,$MeV (a full error analysis is in progress).
These values are very close to those of 
the Gen2 ensemble (also shown in Fig.~\ref{fig:chiral}), which has the
same pion mass and a similar (but
slightly larger) spatial lattice spacing, suggesting that reducing the
temporal lattice spacing alone does not have a big effect on the location of
the chiral transition.

\begin{figure}
\centering
\includegraphics[width=0.45\textwidth]{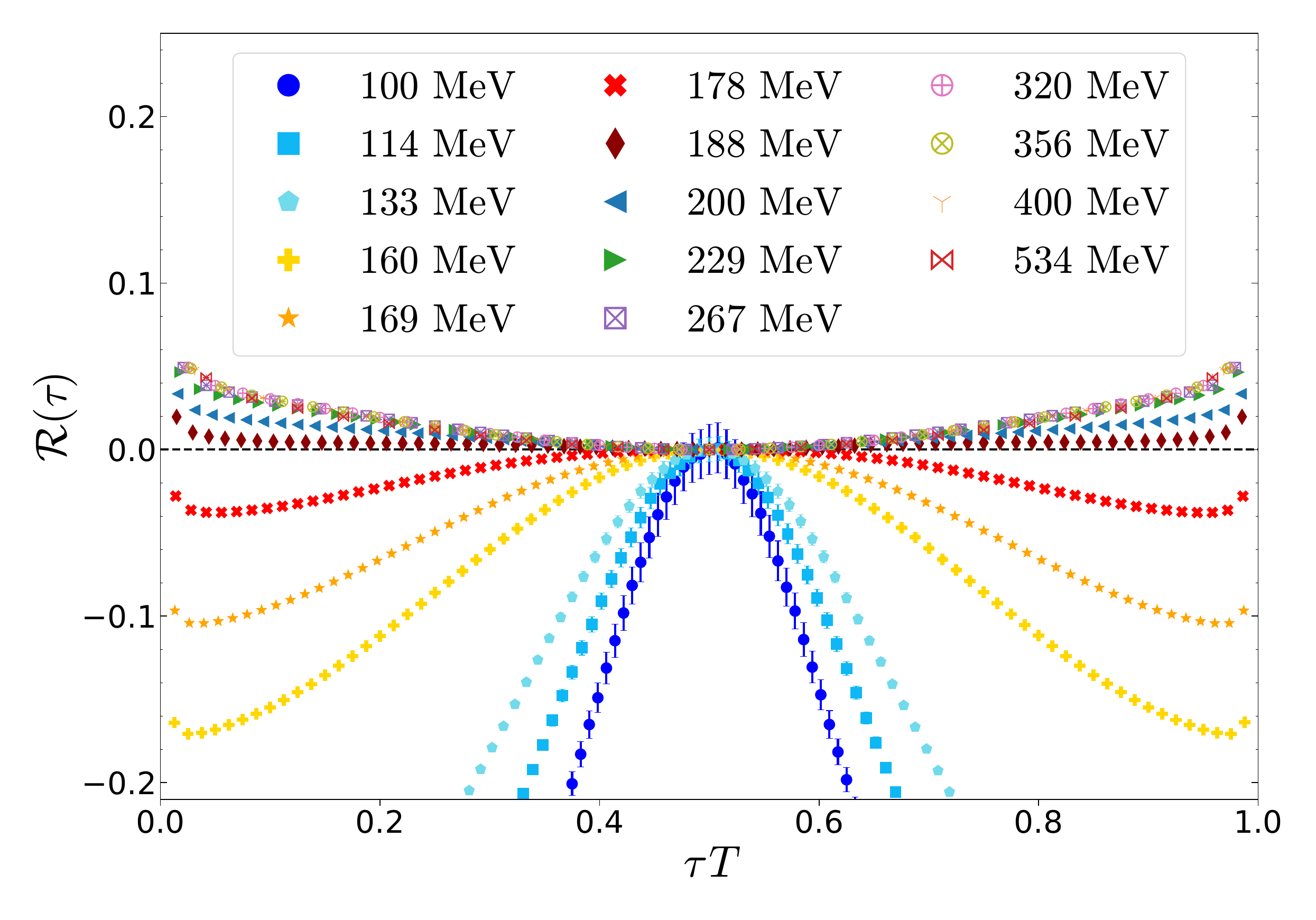}
\includegraphics[width=0.45\textwidth]{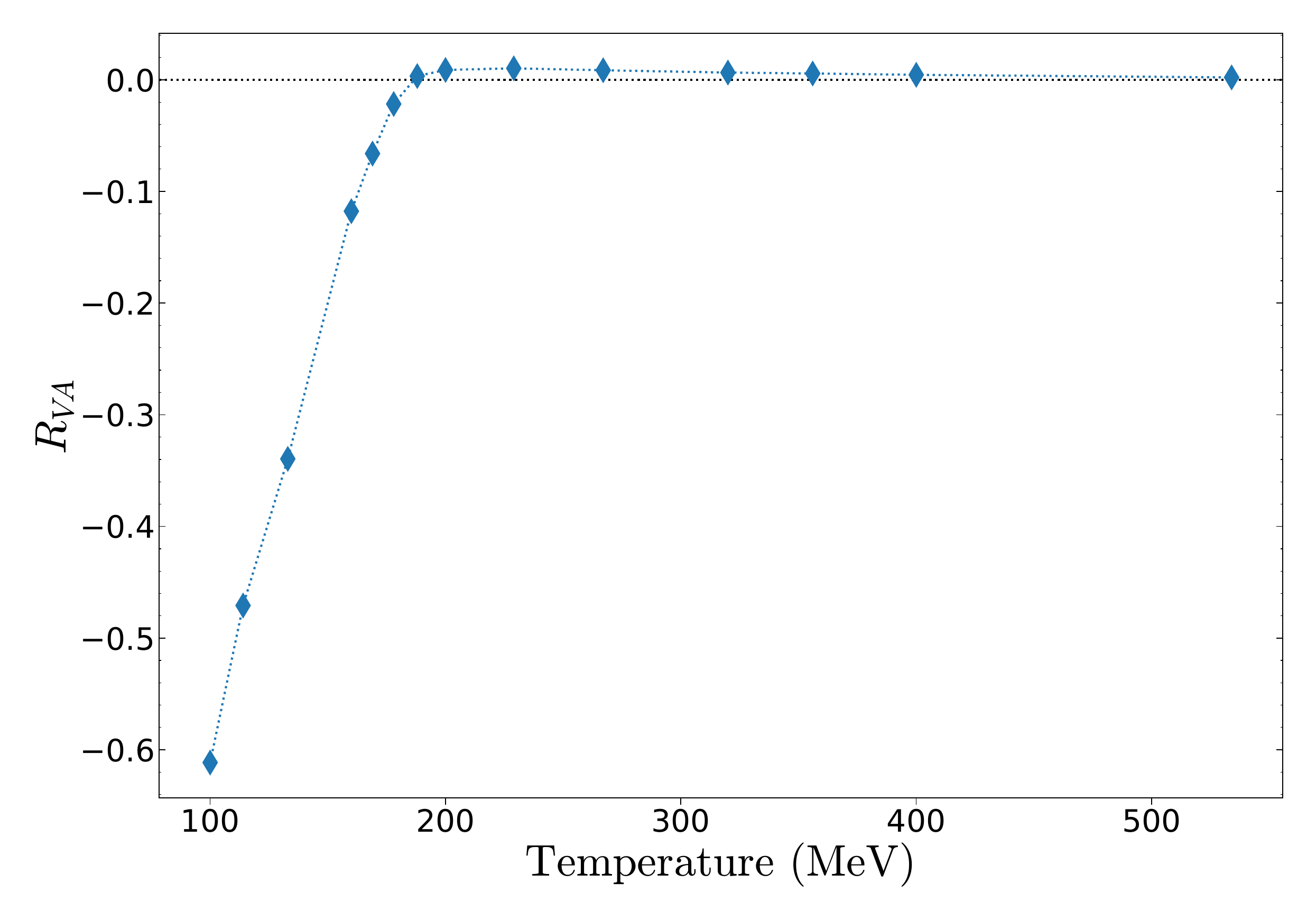}
\caption{Left: The $V-A$ degeneracy ratio $\RVA(\tau)$ for all
  temperatures. Right: The summed degeneracy ratio $\RVA$ as a
  function of temperature.}
\label{fig:V-A}
\end{figure}

Another signal of restoration of chiral symmetry is that the vector
and axial-vector meson correlators become degenerate.  Following
\cite{Smecca:2024gpu}, we introduce the correlator ratio $\RVA(\tau)$
and the summed ratio $\RVA$ as
measures of this degeneracy:
\begin{align}
\RVA(\tau) &= \frac{G_A(\tau)-G_V(\tau)}{G_A(\tau)+G_V(\tau)}\,,
&\RVA &= \frac{\sum\limits_{n=n_{\min}}^{N_\tau/2-1}\RVA(\tau_n)/\sigma^2(\tau_n)}
    {\sum\limits_{n=n_{\min}}^{N_\tau/2-1}1/\sigma^2(\tau_n)}
\label{eq:R-VA}
\end{align}
These are shown in Fig.~\ref{fig:V-A}, using smeared correlators
normalised at $\tau=N_\tau/2$, and taking $n_{\min}=7$.
Since Wilson fermions explicitly break chiral symmetry, $\RVA(\tau)$
will deviate from zero at small $\tau$ even at high temperature;
however, in Fig.~\ref{fig:V-A} we clearly see that the vector and
axial-vector correlators become approximately degenerate above
$T\sim180-200\,$MeV, consistent with the pseudocritical temperature
determined from the chiral condensate.  The pattern is in agreement
with that previously found for the Gen2 and Gen2L ensembles
\cite{Smecca:2024gpu}.

\subsection{Heavy quarkonium}
\label{sec:beauty}

We have simulated beauty quarks on our Gen3 ensemble with a
nonrelativistic QCD (NRQCD) hamiltonian including $\order(v^4)$ terms.
The quark mass was tuned using the dispersion relation for the
$\eta_b$ and $\Upsilon$ states,
\begin{equation}
  E(\vec{p}) = E_0 + \frac{p^2}{2\bar{M}}\,,
  \label{eq:nrqcd-dispersion}
\end{equation}
such that the kinetic mass $\bar{M}$ equals the spin-averaged 1S mass,
$\bar{M}=(m_{\eta_b}+3m_{\Upsilon})/4$.

\begin{figure}
\includegraphics[width=0.45\textwidth]{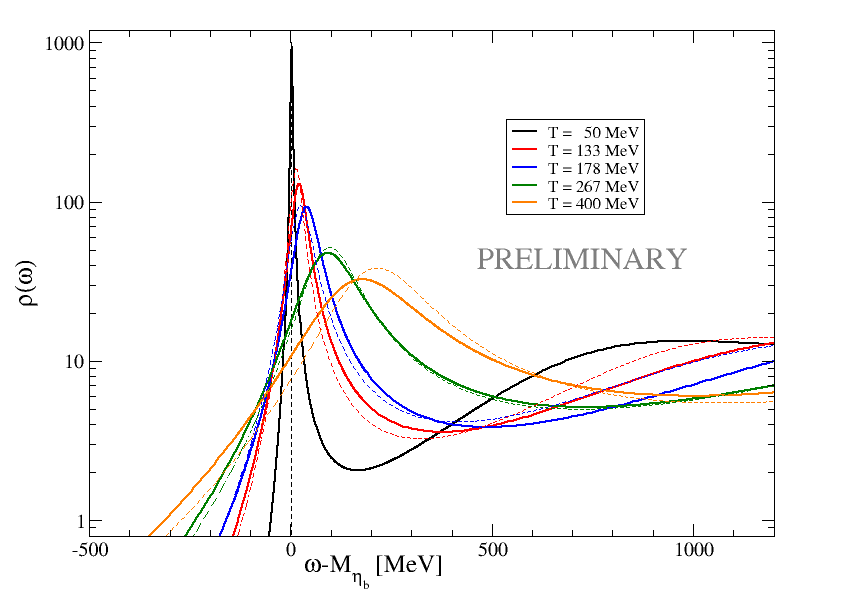}
\includegraphics[width=0.45\textwidth]{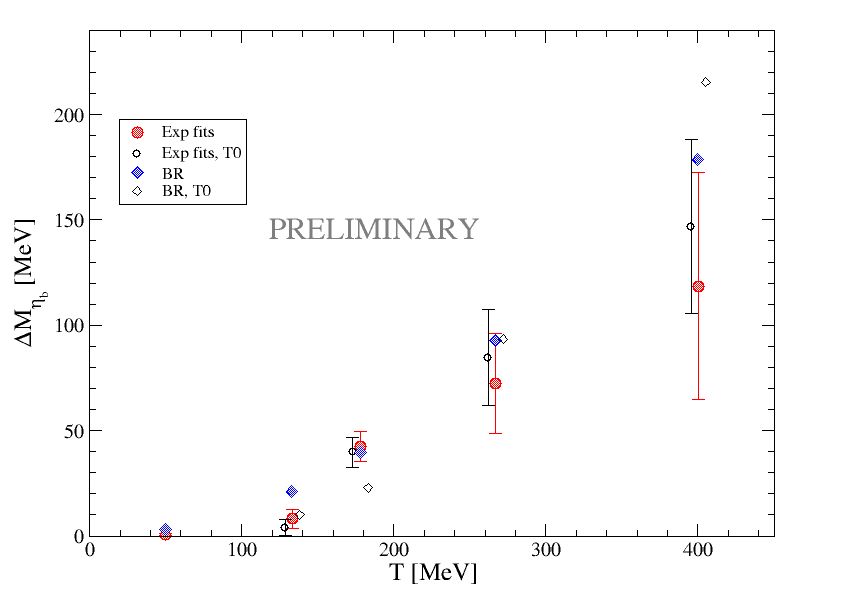}
\caption{Left: $\eta_b$ spectral functions from the BR method at
  different temperatures.  The solid lines are obtained from the
  thermal ensembles, while the dashed lines are obtained from the
  $T=0$ ensembles using the same temporal range (see text).  Right:
  Mass of the $\eta_b$ meson minus its zero-temperature value, from
  exponential fits and from the BR method.  The filled points are
  results from the thermal ensembles, while the open symbols are from
  the $T=0$ analysis with the same temporal range.}
\label{fig:etab}
\end{figure}

Preliminary results for the pseudoscalar ($\eta_b$) state using
exponential correlator fits and spectral function reconstruction using
the BR method \cite{Burnier:2013nla} are shown in Fig.~\ref{fig:etab}.
In addition to the thermal correlators, we have performed the same
analysis on zero-temperature correlators with the temporal extent
limited to $\tau\in(0,1/T)$ for each temperature $T$.  This will
assist in disentangling true thermal effects from those arising merely
from restricting the temporal range in the analysis.  Within the
current uncertainties, these preliminary results suggest that the
apparent mass shift is entirely due to restricting the temporal range
and hence not physical,
but at the present level, a small thermal mass shift as observed in
the Gen2L analysis \cite{Skullerud:2025iqt,Lombardo:2025sfo} is not yet ruled out.

\section{Outlook}
\label{sec:outlook}

We have presented the first results from the new ``Gen3'' ensemble of
anisotropic lattices with a temporal lattice spacing
$a_\tau=15.5\,$am, which is half that of our previous ensembles.  This
is an important step towards controlling lattice spacing effects in our
simulations, and will give improved control in extracting spectral
quantities.  We have shown preliminary results for the chiral
transition including a clear signal of degeneracy between the vector
and axial-vector channels, as well as for the bottomonium mass and
spectral function.  Work on other quantities including light mesons
and baryons and charm physics is underway.

We are also working on tuning parameters for two new ensembles: Gen2P,
which will have roughly the same anisotropy and lattice spacing as
Gen2 and Gen2L, but with physical light quarks ($m_\pi=140\,$MeV), and
Gen3L, which will have the same parameters as Gen3 but with a pion
mass $m_\pi=240\,$MeV.  We expect production of these ensembles to
start shortly, and hence provide further control of the systematics in
our calculations.

\section*{Acknowledgments}

We thank Aoife Kelly for her contribution in the early stages of this work.
G.A., C.A., R.B., T.J.B.\ and A.S.\ are grateful for support via STFC grant ST/T000813/1. M.N.A.\ acknowledges support from The Royal Society Newton International Fellowship. R.B.\ and S.M.R.\ acknowledge support from a Science Foundation Ireland Frontiers for the Future Project award with grant number SFI-21/FFP-P/10186. This work used the DiRAC Extreme Scaling service at the University of Edinburgh, operated by the Edinburgh Parallel Computing Centre and the DiRAC Data Intensive service operated by the University of Leicester IT Services on behalf of the STFC DiRAC HPC Facility (www.dirac.ac.uk). This equipment was funded by BEIS capital funding via STFC capital grants ST/R00238X/1, ST/K000373/1 and ST/R002363/1 and STFC DiRAC Operations grants ST/R001006/1 and ST/R001014/1. DiRAC is part of the UK National e-Infrastructure. We acknowledge the support of the Swansea Academy for Advanced Computing, the Supercomputing Wales project, which is part-funded by the European Regional Development Fund (ERDF) via Welsh Government, and the University of Southern Denmark and ICHEC, Ireland for use of computing facilities. This work was performed using PRACE resources at Cineca (Italy), CEA (France) and Stuttgart (Germany) via grants 2015133079, 2018194714, 2019214714 and 2020214714. We acknowledge EuroHPC Joint Undertaking for awarding the project EHPC-EXT-2023E01-010 access to LUMI-C, Finland.


\bibliographystyle{elsarticle-num}
\bibliography{hot,lattice,jis}
\end{document}